\documentclass[runningheads]{llncs}
\usepackage{graphicx}
\usepackage{cite}
\usepackage{url}
\usepackage[bottom]{footmisc}
\usepackage [autostyle]{csquotes}    
\MakeOuterQuote{"}

\usepackage{lmodern}
\usepackage{textcomp}
\usepackage[utf8]{inputenc}
\usepackage{longtable}
\usepackage[labelfont=bf]{caption}
\newcommand{\tabitemm}{~~~~\llap{\textopenbullet}~~}
\newcommand{\tabitem}{\textbullet~~}
\usepackage{enumitem}
\begin{document}
\title{How to do it right: A framework for biometrics supported border control}
\titlerunning{How to do it right: A framework for biometrics supported border control}
%
\author{Mohamed Abomhara \inst{1} \and
Sule Yildirim Yayilgan \inst{1} \and Anne Hilde Nymoen
\inst{1} \and Marina Shalaginova \inst{1} \and Zoltán Székely \inst{2} \and Ogerta Elezaj\inst{1}}
\authorrunning{Mohamed Abomhara et al.}
%
\institute{Department of Information Security and Communication Technology, Norwegian University of Science and Technology, Norway \\
\email{\{mohamed.abomhara, sule.yildirim, anne.nymoen, marina.shalaginova,ogerta.elezaj\}@ntnu.no} \and
National University of Public Service, Faculty of Law Enforcement, Hungary \\
\email{dr.szekely.zoltan@gmail.com}}
\maketitle              
\begin{abstract}

Complying with the European Union (EU) perspective on human rights goes or should go together with handling ethical, social and legal challenges arising due to the use of biometrics technology as border control technology.   While there is no doubt that the biometrics technology at European borders is a valuable element of border control systems, these technologies lead to issues of fundamental rights and personal privacy, among others. 
This paper discusses various ethical, social and legal challenges arising due to the use of biometrics technology in border control. First, a set of specific challenges and values affected were  identified and then, generic considerations related to mitigation of these issues within a framework is provided. The framework is expected to meet the emergent need for supplying interoperability among multiple information systems used for border control.

\end{abstract}
\section{Introduction}
\label{sec:introduction}
Biometrics technology \cite{jain2000biometric,jain2004introduction} refer to automated methods of identification and verification of the identity of individuals based on their physiological or behavioral attributes.  Examples of biometrics include fingerprints, facial features, iris scans, etc. They are used to support the border police on making decisions by providing  automated identification, verification and cross-checking of individuals based on their biological and behavioral traits \cite{bhatia2013biometrics}.  Identification is a process to associate a person with an identity (who are you?).  Verification is a process to determine whether someone is who he/she claims to be (are you who you claim to be?). Cross-checking is a process of verifying information by using alternative European Union (EU) information systems.   Biometrics technology is increasingly being used by countries worldwide and is a highly adopted technology at the EU borders \cite{diaz2014legal,tanwar2019ethical}. It aim is to help achieving an automated, rapid and highly secure border clearance process, such that an increasing passenger throughput does not compromise border control reliability.  

On the one hand, biometrics technology has been proven to be cost-effective to enhance border security, detect fraud and help to improve border crossing efficiency as well as facilitate an effective migration control and enforcement. On the other hand, biometrics technology can lead to some challenges and conflicts with fundamental human rights and can be a cause of ethical, social and legal challenges \cite{de2013biometrics,tanwar2019ethical}.  The key challenge is related to individual rights, such as respect for personal privacy \cite{zeadally2015privacy,yu2016big}, human dignity \cite{floridi2016human}, bodily integrity \cite{van2007genetics}, equity and personal liberty \cite{sutrop2010ethical,wickins2007ethics}. Personal data protection is also an issue, especially when biometrics information is stored in centralized databases \cite{sprokkereef2007data,campisi2013security}.  

Another major concern with biometrics technology is the seemingly immutable link between biometric traits and persistent personal information storage about individuals \cite{national2010biometric}. 
The tight link between personal records and biometrics can have both positive and negative consequences for individuals and the society overall. Recent research \cite{dantcheva2015else} on biometrics data shows that it can reveal personal information, such as gender, age, ethnicity and even critical health problems like diabetes, vision problems, Alzheimer's disease, etc. Such confidential information might be used for example to discriminate among individuals when it comes to border crossing enforcement.

People have the right to choose to what extent and how to engage with the systems and devices (e.g., biometric sensors). For example, some people may refuse to have their photographs taken by a face recognition system due to the concerns about the purpose of the use of the images \cite{national2010biometric}. Moreover, others may refuse to or feel uncomfortable to undergo iris scans or provide fingerprints due to permanent or temporary disability. A study of biometric enrollment and verification in the United Kingdom
showed that 0.62\% of the sample group of people with disabilities were unable to enroll any of the three biometrics tested: fingerprints, facial scans, and iris scans \cite{lee2016biometrics}.   Moreover, according to the European health and social integration survey (EHSIS), in 2012 there were 70 million people with disabilities in EU \cite{Disability}. If 0.62\% of all disabled in EU were unable to provide this data, approximately 434000 Europeans would be unable to provide any of the three most common types of biometric data. Such concerns may impact these people and also people belonging to different groups with varying cultural beliefs, values and specific behaviors on how they interpret the requirements of being exposed to biometrics technologies. In general, a range of complex and interconnected issues must be addressed while deciding on the use of biometrics as a technology for border control \cite{sprokkereef2007ethical}. Also, ethics guidelines and a regulatory framework for the use of biometrics technology in border control must be formulated in order to avoid  any harmful impact on the society while allowing for the continuous development of this technology to benefit the society \cite{de2013biometrics}.

In this paper, section \ref{sec:background} discusses the potential benefits of using biometrics in EU borders and the ethical theories around it. Section \ref{sec:DoItRight} investigates the key ethical, social and legal challenges of using biometrics technology and demonstrates vulnerabilities and risks related to these challenge categories, followed by a discussion of  moral considerations with regards to human rights, right to privacy, right to data protection etc. Section \ref{sec:DC} presents a discussion on ethical reasoning and decision making. Section \ref{sec:CONclud} concludes the study.


\section{Background}
\label{sec:background}
This section provides a background of benefits of biometrics in EU borders and a discussion on moral, ethics and ethical theories.

\subsection{Potential benefits of using biometrics in EU borders}
\label{sec:benefits}

Freedom of movement is restricted by closed or controlled borders in order to protect other fundamental rights such as security or health, national or regional political, societal, cultural or economic interests of the political entities within that bordered area. 
The Schengen Border Code (SBC) (Regulation (EU) 2016/399) and its amendment (Regulation (EU) 2017/458) set out the rules governing the movement of people across EU's internal and external borders. The main aim of border checks is to ensure that the persons and goods crossing the border are entering or leaving the area with the permission (authorization) of the political entity of SBC.  This permission for travel is currently manifested in a travel document having a physical form as well as a record on the authorization in the national travel document database.


Identification and verification procedures at the border are to ensure that  entry-to or exit-from a country will be granted to the right persons. In recent years, Member States have seen an increased use of biometric identification and authentication systems at EU's borders including airports and land borders \cite{labati2016biometric,anand2016enhancing}. More significantly, the large-scale EU information systems such as Visa Information System (VIS), Second-generation Schengen Information System (SIS II), European Asylum Dactyloscopy Database (EURODAC) and Entry Exit System (EES) etc. \cite{kenk2013smart} employ biometrics for migration and border control and management. Such systems involve several highly complex processes, leading to a number of ethics and privacy challenges \cite{sutrop2010ethical,de2013biometrics}. The integration of biometric at border control provides benefits for travelers, political entities (states),  authorities responsible for border control as well as individual border guards. Such benefits include accuracy, integrity, robustness and efficiency. 


\begin{itemize}
	\item \textbf{Accuracy:} Accuracy of travelers' identification and verification means the ability to recognize genies person and reject imposters person correctly \cite{ICAO2017}. During manual border checks, border guards seek to gain knowledge about the subject (traveler) and associate it with his/her identity.  
	For example, the border guard looks at the traveler and verifies with the picture on the travel document (on bio-data page) to determine if the person standing in front of him/her is the same that is pictured in the travel document. However, the accuracy of identification and verification depends on lighting, age of the picture, perception capabilities, tiredness, make-up etc. In addition, spreading culture of having aesthetic surgery poses a further challenge to manual identification. Moreover, a human border guard is usually very efficient during the start of the shift, then diminishing attention appears as the officer gets tired. In this case, biometrics can enhance and support these practices. A selective and differentiated application of multimodal biometrics identification results in a higher average accuracy during the whole shift as well as  facilitates cross-checking of personal data with greater accuracy \cite{ICAO2017}. 
	
	
	\item \textbf{Integrity:} Integrity of the identification is the ability to confirm that the collected data and its components (e.g., passport picture and passport information) have not been altered from that created by the issuing State or organization \cite{ICAO2017}. Use of biometrics enhances the reduction of identity fraud impersonation ( e.g., fake IDs and passports) as the identification and verification processes do not rely on the human agent. 
	To the best of our knowledge, a reduction in fraud means an increase in accuracy. Therefore,  using biometrics eliminates a quite considerable integrity threat that the border guards face and benefits the authority responsible for border control.

	
	\item \textbf{Robustness:} Biometrics systems are easy to operate, maintain, update, replace, redeploy or decommission compared to border control units/booths consisting of human agents only. Long years to achieve full competence and repetitive training is not required and experience is not lost with a single unit. From the traveler's view, utilizing multimodal biometrics allows the traveler to (theoretically) decide which biometric modality (e.g., fingerprints, face, iris) will be used for identification based on his/her preferences. 
	
	\item \textbf{Efficiency:} The processing capacity of Automated Border Control (ABC) gates is sustained over time as ABCs don't get tired. Additionally, ABCs conduct an objective repeatable set of checks to complete identity and document verification can be more accurate and quicker to complete than similar checks conducted by humans \cite{ICAO2017}.  This results in a higher number of low-risk travelers' throughput without losing accuracy or integrity and allows human resources to be focused on potentially higher-risk travelers.

\end{itemize}

As these benefits are all potential, the actual benefits highly depend on how biometric systems are integrated into the border management and how control systems help facilitate the correct identification and verification of persons and contribute to fighting identity fraud. 

\subsection{Moral, ethics and code of ethics}
\label{sec:Ethical}

Morals are the general views, thoughts and convictions of people in making judgments about what is right or wrong. According to Kizza \cite{kizza2007ethical},  morality is defined as "a set of rules (code) of conduct that governs human behavior in matters of right and wrong, good and bad."  Ethics, on the other hand,  concerns the way we can come to moral judgments of what is right and wrong for individuals and society.  The ethical judgment of what is good or bad and right or wrong is often based on a set of shared rules, principles, and duties applicable to all in a group or society and this is called code of ethics. A code of ethics is a written set of ethical principles and guidelines that govern decisions and behaviors in an organization (e.g., border management authorities) according to its primary values and ethical standards and  theories \cite{boddington2017towards,banks2018criminal}. 

There are many ethical theories, each of which emphasizes different points, such as predicting an outcome and carrying out one's duties to others to reach an ethically correct decision. Consequentialism theory (result-based ethics) emphasizes the consequences of human actions, whether good or bad, right or wrong. Deontological ethics does not concern the consequences of an action but rather it considers the will and the motivation for undertaking an action \cite{kagan2018normative,ronzoni2010teleology,kizza2007ethical}. It is sometimes described as duty-based or rule-based ethics.  
Even though the distinction between deontological and consequentialism is often clear, the two theories are fundamentally different. They are both normative, and as a result, code of ethics are formulated as guidelines rather than prescriptions and prohibitions. The aim of the code of ethics is to provide a moral basis for emerging professional choices and provide adequate protection for all those who act in a statutory manner, and to recognize the unworthy practices associated with the police profession. The following are examples of the ethical principles for border guards which provide moral guidance during service shifts and out of service shifts.

\begin{enumerate}
	\item \textbf{Respect of dignity and rights}: Respecting the rights of every person and avoid the use of torture, inhuman or degrading treatment.
	\item \textbf{Fairness}:  Treating of persons must be equal and having no preference, bias or prejudices based on race, background, ethnicity, gender, religion, personal and social status or property status etc.
	\item  \textbf{Duty of confidentiality}: Respecting for privacy and guaranteeing the security of the data and the information obtained.
	\item \textbf{Responsibility}: Taking responsibility for actions and decisions in legal and moral terms. If misconduct takes place, take steps to ensure it is not repeated.
\end{enumerate}

\section{The "how to do it right" framework}
\label{sec:DoItRight}
\begin{figure}[t]
\includegraphics[width=\textwidth]{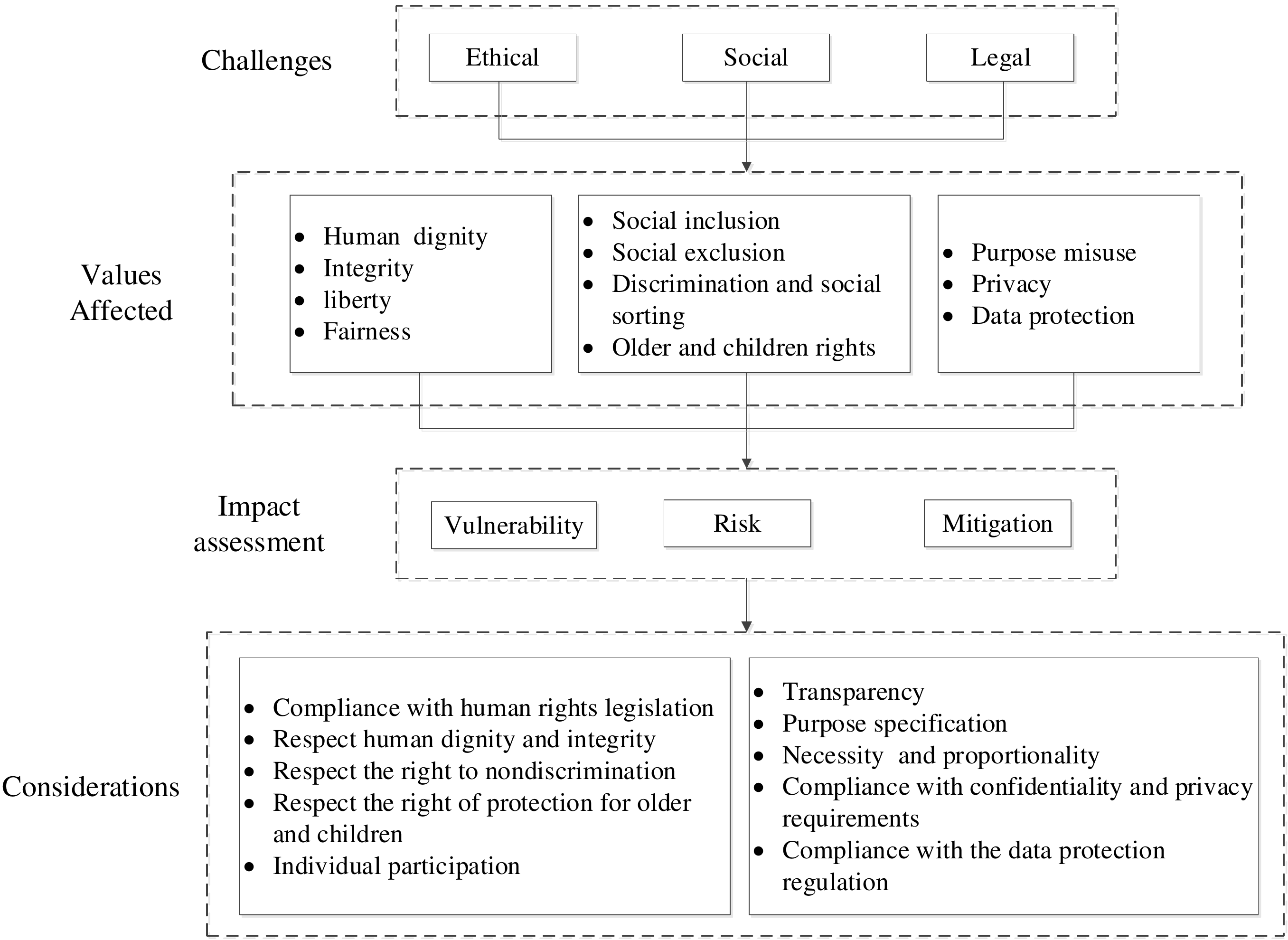}
\caption{The proposed framework for specifying biometric technologies related challenges and the considerations} 
\label{fig1}
\end{figure}

 This section presents a framework (Figure \ref{fig1}) which helps to point out the types of challenges, the specific vulnerabilities and risks which the stated concerns/challenges lead to and the generic considerations for handling these challenges raising due to the use of biometric technologies. The focus on ethical, social and legal challenges constitute the topmost layer of the framework. The next layer of the framework is the values affected due to the presence or uprising of the challenges. The third layer from the top is the impact assessment layer. That is, what are the consequences of a value being affected? What vulnerabilities and risks arise correspondingly and what are the mitigation plans? Then, the bottom most layer lists the corresponding considerations that the border control police are expected to comply with.  The framework provides a what to do and how to do it right guideline for border control police when various challenges are met by providing a link from the challenge, to the value(s) affected and  an impact assessment on  the values affected as well as pointing out the considerations that must be in place. Below, we provide  examples of which values maybe affected under varying challenges and how.


\subsection{Challenges level}
Biometrics technology is acknowledged to potentially raise critical ethical, social and legal challenges. Ethical challenges generate situations that requires a person or organization to choose between alternatives that must be evaluated as right (ethical) or wrong (unethical). Social challenges create problems that engaging in normal social behaviors and may influence a large number of individuals within a society. Legal challenges refer to formal questioning of the legality of an act and whether or not the act is being taken in accordance with the law.

Challenges layer provides information to help border officer or border authorities think through basic ethical, social and legal concepts and considerations. It will not provide specific answers for the specific challenges but will help bring to conscious awareness some understandings that help in thinking through these issues.

\subsection{Values level and examples of how values are affected by ethics, social and legal challenges}
\label{sec:ethical}

Below, we give examples of how several values may be affected due to the arisen of such challenges.


\begin{enumerate}[label=(\Roman*)]


\item \textbf{Human dignity:} The capability to verify travelers' identities is extremely important and regards human dignity \cite{sutrop2010ethical,floridi2016human,kizza2007ethical}. People may feel uncomfortable (or humiliated to some extent) when authorities like the border police are recording body features algorithmically. Many factors, for example physical work or physical incapacity (e.g., physical disabilities, sight impairment and mental health problems), can make it hard for some people to provide biometrics or they may simply be unwilling to do so. For instance, damaged fingers due to manual work can impact the way people are treated when providing fingerprint data \cite{drahansky2012influence}. In these cases, challenges with collecting biometrics data and remaining respectful of human dignity may emerge. People who cannot provide fingerprints or other biometrics data sometimes face a greater risk of negative consequences than people who can \cite{lee2016biometrics}. 

Biometrics data collection from vulnerable persons including those with disabilities requires particular attention. Human dignity is evidently a complex notion of the individual and biometrics data is strictly linked to the human body, whose integrity (physical and psychological) constitutes a key element of human dignity.  


\item \textbf{Social inclusion/exclusion and risk of discrimination:} The introduction of biometrics to improve identity verification at land borders raises serious objections to the potential to facilitate discriminatory social profiling \cite{wickins2007ethics}. For example, the biometrics enrollment of injured and disabled travelers could lead to higher false rejection rates than average. Moreover, senior citizens and children who have particular problems with using biometrics enrollment devices (e.g., fingerprint scanner, iris recognition reader) may face enrollment difficulties. Although discrimination of vulnerable individuals might be involuntary and unintentional, it may deeply affect them and impact the principle of equity. Furthermore, religious aspects (e.g., beard, headscarf) or interpersonal contact (e.g., photographs, touching, exposing parts of the body) may render a biometrics system an unacceptable intrusion. For example, those of faith who wear head or face coverings have difficulties with enrolling facial biometrics. Verification of such biometrics in public (e.g., at the border crossing points) may lead to embarrassment or offense, causing avoidance of situations where this is necessary. Therefore, mandatory encouraged use of such system may undermine religious authority and create de facto discrimination against certain groups whose members are not allowed to travel freely or obtain certain services without violating their religious beliefs and privacy. In positive terms, respecting a person's intrinsic worth requires recognizing that the person is always entitled to participate in social and community life regardless of age, beliefs, disability, health, etc.


\item \textbf{Children rights:} Biometric technology also present several ethical questions regarding children's rights. These include the right to information, the right to privacy, security and the right to no discrimination, etc.  With respect to children and  biometric technology, the main concern is that children will not fully know or understand the implications of the accessibility to, and subsequent use of the data collected. While children (and indeed their parents) may be aware of basic privacy settings and risks, even sophisticated users face great difficulties. Moreover, child identification introduces the requirement for greater levels of care. The problem is that there are several reasons why not all biometrics can be used for child identification and biometric recognition of toddlers.  For example, a study by Basak et al. \cite{basak2017multimodal} found that "capturing fingerprints for children less than three years is hard due to very small fingerprint area, smooth skin, and thin fingers." Therefore, very young kids with small fingerprints might not be identified efficiently.

Furthermore, children are particularly entitled to effective privacy protection. This is because children cannot develop privacy expectations for reasonable legal protection. 
Moreover, biometric match accuracy diminishes as children grow. Fingerprinting young children affects the quality and reliability of future matches to the initial fingerprints \cite{jain2017ingerprint}. The risk of a wrong match increases when the fingerprints or facial images are compared more than five years after the initial collection. 


\item \textbf{Purpose misuse:} Function or purpose creep occurs when the biometrics data is collected for one specific purpose and subsequently used for another unintended or unauthorized purpose without the user's consent. A famous example of a large-scale biometric function creep is the European Dactyloscopy (EURODAC) fingerprint database. The original purpose of this EURODAC was to compare fingerprints for the effective application of the Dublin convention. It enables EU countries to identify asylum applicants as well as illegal immigrants within the EU. However, soon after the database was established, other police and law enforcement agencies were also granted access.  Similar concerns may also arise in the case of other large-scale, centralized EU national and international databases, such as SIS II, VIS and EES. Biometrics are likely to strengthen the potential for function creep due to the very sensitive nature of the data collected and the possibility to use centrally stored biometric data for purposes other than the original purpose.


\item \textbf{Right to privacy and data protection:} Every individual has the right to privacy protection and personal data protection when his/her data is collected and shared. The use of biometrics technology as a border control tools introduces problems with maintaining individuals' privacy and protection of their personal data. Such a technology will probably increase the risk of available information misuse as a result of unethical and/or illegal practices if personal data  are not protected adequately.

The main concerns include unnecessary and unauthorized collection of biometrics data for traveler identification and verification \cite{campisi2013security,zeadally2015privacy}.  GDPR \cite{voigt2017eu}, among other legislation, state that to best preserve an individual's privacy and right for data protection, the amount of personal data collected should always be kept to a minimum. Moreover, personal data like biometrics data should only be used when individuals or authorities will benefit from the collection. 
Cameras, for instance, are now widely used to monitor our everyday life. People often benefit from such monitoring, especially at borders to control people flows and detect suspicious activities (e.g., illegal border crossing). However, extensive data collection and analysis can also lead to privacy violations.  

Information linkage and compromise of anonymity is another concern \cite{zeadally2015privacy}.  Various kinds of information about individuals stored in a range of databases (e.g., SIS II and VIS) have the potential to become yet another means through which information can be linked to purposes ranging from commercial marketing to law enforcement.  Recent research \cite{dantcheva2015else} explores the possibility of extracting supplementary information from primary biometric traits, face, fingerprints, hand geometry and the iris. Such information includes personal attributes like gender, age, ethnicity, hair color, height, weight and so on.

Despite all the benefits of using biometrics technology in border control, privacy concerns have become widespread because each time a person's biometric data is checked, a trace is left that could reveal personal and confidential information. 
Biometric data should essentially be well-protected against unnecessary and unauthorized collection, access and  disclosure etc.. 

\end{enumerate}


\label{sec:framework}


\subsection{Impact assessment level}
\label{sec:impact}


Table \ref{impactTable} summarizes values that may be affected and maps them to vulnerabilities, risks and possible mitigation measures in comparison to the current systems authorized in EU border control and other solutions already on the market.

\begin{longtable}{|p{1.8cm}|p{2.9cm}| p{2.7cm}|p{4.3cm}|}
\caption{Values and the corresponding  vulnerabilities, risks and mitigation measures.\label{impactTable}}\\
\hline
\textbf{Values} &  \textbf{Vulnerability} & \textbf{Risk} & \textbf{Potential mitigation measure}\\
\hline
\endfirsthead
\hline
\textbf{Values} &  \textbf{Vulnerability} & \textbf{Risk} & \textbf{Potential mitigation measure}\\
\hline
\endhead 
\endfoot
\hline
\endlastfoot
 Respect to human dignity & Current systems do not afford individual a choice of what biometrics data they prefer to enroll or use. &  Violation of right to human dignity, cultural or religious customs etc.   &   Border control biometrics must provide information about what and why biometrics used  as well as allow choice policies and procedures, unless choice is inapplicable. \\

\hline
 Right to the person integrity &  Current systems do not adapt or lack informed consent policies.   & Violation of the right to integrity.  &  Border control biometrics must ensures a collection of free and informed consent form individuals according to rules laid down by regulations such as GDPR. \\

\hline
 Right to person liberty &  Current systems may lack of policy, procedures and ethical guidelines for data collection and processing or may allow unauthorized processing of individuals data; e.g., use of force to collect biometrics data.     &  Violation of the right to liberty of an individuals.    &   Border control biometrics must apply policy to restrict the procedures of data collection and processing, balancing between lawful interest and personal liberty.   \\

\hline
 Right of protection for children &  Current systems does not address/deal with children vulnerability and special needs   &  Violation of the children rights which may lead to high levels of discrimination due to children lack of knowledge about the systems.  &  Border control biometrics must envisages adoption of devices and procedures to ensure children' needs. Also, the biometric data of children should be treated with enormous care and the procedures need to comply with data protection legislation such as GDPR \cite{voigt2017eu}. Parents must always be notified when their children's biometric data is to be collected or used, and written consent must be obtained in advance. \\

\hline
 Right to no discrimination &  Current systems may be discriminated based on sex, race, ethnic or social origin, genetic features, religion or belief, political opinion, disability, age or sexual orientation etc.  &  Discrimination, social inclusion/exclusion and social sorting of individuals. &   Border control biometrics must ensure nondiscrimination policy that comply with human rights legislation. \\

\hline
Respect for private and family life &  Current systems do not adapt an adequate family related consent and procedures.   &  Violation of right to respect for private and family life.  &  Border control biometrics must adopt appropriate measures for family consent and procedures. \\

\hline
 Right to no information tracking &  Current systems lack of notices and information about tracking of individuals. &   Violation of personal right and legitimate purpose requirements leading to surveillance of individuals and/or other members of the family.    & If within the purpose and the law, surveillance should be authorized and consistent with EU and national laws.   \\

\hline
Right to personal data protection &  Current systems may allow personal data to fall into the wrong hands or/and shared across organizations.   &  Violation of right to security principles such as confidentiality, integrity, and availability.  & Border control biometrics must ensure Security by Design to guarantee the ongoing confidentiality, integrity, availability and resilience of processing systems and services.   \\

\hline
Right to privacy and confidentiality &  Current systems may breach confidentiality, allowing unauthorized disclosure of personal information.   &  Violation of personal privacy.    &  Border control biometrics must ensures implementation of privacy-enhancing technologies to protect data in accordance with the law. \\

\hline
\end{longtable}

\subsection{Considerations Level}
\label{sec:consd}
This section presnets ethical, social and legal considerations. 

\begin{enumerate}

	\item \textbf{Ethical considerations for human rights:}  According to Article 7, Regulation (EU) 2016/399 (amended in Regulation (EU) 2017/458), competent authorities should ensure that the human dignity and integrity of persons whose data are requested are respected and should not discriminate against persons on grounds of sex,religion, disability, age etc. Thus, biometric platform at border should be designed to support human right-compliant systems, related to technological, ethical and sociological aspects. For biometric technologies to be successful with its use and actual implementation, they should not only consider the security and privacy of personal data, but it also need to guarantee that the users can interact with the systems and make the user experience acceptable. To do so, system designers and policy makers must consider all challenges, vulnerabilities and risks (Table \ref{impactTable}) related to the system design. Moreover, border control biometrics platforms must pay particular attention to minors, whether traveling accompanied or unaccompanied and must respect the specific needs of children and their interests must be protected in ways supplementary to the general treatment of adult subjects.

\item \textbf{Considerations for travelers with physical or mental impairment:} As mentioned above, EU Regulation (EU) 2016/399 specifies equal rights for border crossing. 
Therefore, border control biometrics platforms should consider travelers with special needs/categories including individuals who physically or mentally impaired etc. An extra attention should be given to, among many others: 
\begin{enumerate}
	\item People with temporary injuries who might have difficulties to provide biometric sample due to temporary wound (e.g., injured face and/or broken arm/fingers) \cite{lee2016biometrics}. In this case, Border control biometrics platforms should not discriminate against such people and shall use biometric devices (e.g., fingerprints scanner) which perform acquisition in a greater number of situations.
	\item People with total permanent disability whom have difficulties to freely move their limbs due to sensory damage and/or muscle damage.  For example, in case of fingerprints verification, travelers with a hand disability may lack the ability to place the required finger and keep it steady for a sufficient time on the fingerprints scanner. Moreover, in case of face recognition/iris scanning, people with neck disabilities may have difficulties in correctly placing their face near the iris scanning device/face recognition camera. Thus, border control biometrics devices should be able to work in off-axis acquisitions and be adjustable to support such people with biometrics recognition and make it more comfort. 
\item People with technological illiteracy, for example, elderly people and children who lack knowledge of using technology/tools (e.g., automated border control gates) would have a difficulty to use and interact with devices. In this case,  border control biometrics devices should design an interface that taking into consideration elder's and kids' needs.
\end{enumerate}
  
	
\item \textbf{Considerations for privacy and data protection:} As mentioned earlier, 
biometric data can be used to recognize individuals automatically with greater accuracy. On the other hand, a misuse of such biometric data can have dangerous consequences which pose several security and privacy challenges such data destruction and/or unauthorized disclosure of, or access to personal data, to name a few.  Thus, border control biometrics platforms should be designed to support privacy-compliant biometric systems.  Perceived risks are related to how people view the biometric technology, whether they trust it, and whether they like to use it. With respect to this, EU regulation (e.g., GDPR \cite{voigt2017eu}) prohibit the use of special categories personal data such as biometric data without the user's awareness and permission. Also, prohibit the use of the biometric data different from the purpose of the system (purpose misuse issue discussed in section \ref{sec:ethical}). For example, biometric data stored in e-passports can only be used for issuing electronic documents and verification of document holder (Regulation (EC) 444/2009).  

Border control biometrics platforms should consider several privacy aspects for protecting the privacy of personal data. These aspects include: 
\begin{enumerate}
	\item The purpose of biometric data: The legitimate purpose of biometric data collection and processing used only for verifying the identity of the individual during the border crossing procedure.  Article. 13 (1) of the GDPR stipulates that "information to be provided to data subject where personal data are collected from the data subject."  This information shall include, purpose of the system, the enrolment and verification processes, and the methods used for data protection, among other.
	\item People control of their personal data: According to Article. 32 (2) of the GDPR \cite{voigt2017eu}, the data subject has the right to ask for removal or erasure of biometric data in electronic documents. Also, the data subject should have the possibility to decide when he/she no longer be authenticated and verified using the biometrics system and choice to proceed with manual checks (when applicable).
	\item Data protection measures:  Article. 32 (2) of the GDPR stipulates that \textit{"the controller and processor must implement appropriate technical and organizational measures to protect personal data against destruction, loss, alteration, unauthorized disclosure of, or access to personal data transmitted, stored or otherwise processed and against all other unlawful forms of processing."} Therefore, border control biometrics system shall deploy a privacy enhancing technologies and secure access control techniques to avoid any misuse of personal data. 
	\item Reuse of data for law-enforcement purposes: According to Regulation (EU) 2016/399 and its amendment Regulation (EU) 2017/458, citizens should be checked in criminal databases such as SIS II  and SLTD on a systematic and non-systematic basis. As the majority of these travelers are presumably innocent individuals. Therefore, saving and cross checking their data on a systematic basis with law-enforcement databases would be disproportionate. Any reuse of personal data done for the purpose of law enforcement should be done in accordance with Directive (EU) 2016/680 \cite{Dir2016-680}, which aims to ensure more consistent and higher level of protection of the personal data of natural persons in the areas of criminal matters and public security.
	
\end{enumerate}

\end{enumerate}

\section{Discussion on ethical reasoning and decision making}
\label{sec:DC}


In view of the ethical theories (discussed in section \ref{sec:Ethical}) and the open dilemma of what is right and what is wrong, it is clear that the situation is similar, particularly surrounding the use of biometric technology. On the one hand, one group of people (travelers, border officers etc.) may see biometrics technology used in border control as a liberator, believing in the power of technology to bring convenience (e.g., avoid queues) and efficiency (e.g., cut costs) and increase mobility (e.g., convenient border-crossing for citizens). This group may also welcome more powerful surveillance to improve border security (e.g., monitor migration, combat identity theft and fraud etc).   We may agree with this group. First and foremost, the use of biometrics in the border control aims to improve security and detect fraud (discussed in section \ref{sec:benefits}) which will leads to maximizing the benefits for the society (e.g., travelers) and minimizing the human workload (e.g., border police officers). With respect to the consequentialism theory (discussed in section \ref{sec:Ethical}), every society member (travelers, border police officers etc.) must benefit the same and it is not specific to any individual. Furthermore, the reason one individual must promote the overall good is the same reason why anyone else has to promote the good. Hence, it can be said that the ethics of the border control biometrics is related to consequentialism. The consequentialism theory places a group's interest and happiness above those of an individual for the good of many.

On the other hand, other groups of people may object and perceive biometrics technology as a threat to their personal life and privacy. Such groups might believe that surveillance technology is untrustworthy and destructive to liberty, dignity and privacy. For example, collecting biometric data such as iris scanning from veiled Muslim women \cite{rahman2018biometrics} in stressful situations (e.g., inappropriate police behavior due to exhaustion or stress) may undermine the dignity of the women being scanned.  An FRA report "Under watchful eyes: Biometrics, EU IT systems and fundamental rights" \cite{EU2018} showed that disproportionate force has been used when fingerprinting asylum seekers and migrants in irregular situations.  Considering deontological ethics (duty-based or rule-based ethics discussed in section \ref{sec:Ethical}) and given the vulnerability of the people concerned as well as the obligation to use the least invasive means, it is difficult to justify the use of physical or psychological force solely to obtain biometrics for the purpose of identification and verification. When it comes to border control and border rules, ethical theories might change according to circumstance. Border officers have a duty to do the right thing (verify the identity of a traveler before entering/exiting the border etc.) even if it produces an undesirable outcome. In the case of veiled Muslim women or any other cases, it would be difficult to judge the action of an officer based on the outcome.

From the review, it could be concluded that ethics are not absolute, and clearly, views on biometrics technology vary according to the differing needs of people and institutions. However, different perceptions of biometrics technology reflect the diverse value judgments as influenced by many factors: age, gender, cultural beliefs, education, moral imagination etc.  Remarks over the use of biometrics technology for large populations, especially if the consequences lead to social exclusion, either as a result of the individual being unable to reliably enroll or verify their data, or simply not having confidence in the system and avoiding having to interact with it. Certainly, when it comes to border control and the use of biometrics technology to increase border security, monitor migration and combat identity theft and fraud etc.,  the argument is essentially utilitarian (consequentialism theory) where the collective right of a group (group interest) is balanced against the rights of the individual. It makes the individual simply a means to the ends of the majority. However, this could be a wrong argument. Wickins in \cite{wickins2007ethics} and Townend in \cite{townend2017overriding} argue that public interest must be judged by considering the balance between individuals, i.e. the rights of single individuals must be balanced against other single individuals if individuals are not to be used instrumentally.

\section{Conclusions}
\label{sec:CONclud}
An important conclusion to this paper is that we are not attempting to provide an answer to what is ethical and what is not, or what is right and what is not. We see biometrics technology usage in border control with two sides. One side is the main intention and aim of biometrics technology to improve border control management and enhance people flow etc. The other side represents the risk of violating personal rights. As said, conflicts with decisions based on what to choose (e.g., privacy versus security, autonomy versus solidarity) make it difficult to have a broad and consistent position in favor of, or against expanding or restricting biometric technologies.

Individual acceptance of biometrics technology should be actively promoted through ensuring transparency of decision-making, clear policy regarding the purpose of biometric technology and how it is used, as well as increased measures dedicated to preserve personal rights and personal data protection. Since greater use of personal data impacts upon human rights, there needs to be an honest and assertive study of what the risks are to personal rights and privacy as well as how these risks are mitigated. Border control biometrics should comply with human rights legislation to encourage respect for fundamental rights in the implementation of biometrics technologies. Also, they should respect human dignity and protect personal integrity, preserve individual freedom and self-determination (i.e., choice and consent with respect to which biometrics data he/she prefers to use), respect privacy and family life, and safeguard against harm and unreasonable force for data processing. Moreover, border control biometrics should comply with security requirements and data protection legislation to ensure data confidentiality, integrity and availability when collecting and processing personal related data.

In the future, we shall investigate the use of ontologies for knowledge representation and enhancement of knowledge discovering using machine learning techniques.  Ontologies provide a formal, explicit specification of a shared conceptualization of a domain that can be communicated between people and heterogeneous and widely spread application systems \cite{taye2010understanding}. We aim to propose a semantic based framework for biometrics integration in border control systems relying on ontologies and machine learning techniques \cite{buitelaar2005ontology} to tackle ethical, social and legal  challenges.

\section*{\uppercase{Acknowledgements}}
This work is carried out in the EU-funded project SMILE (Project ID: 740931), [H2020-DS-2016-2017] SEC-14-BES-2016 towards reducing the cost of technologies in land border security applications.

%
%
%
\bibliographystyle{splncs04}
 \bibliography{Bibliography}

\end{document}